\documentclass{aa}

\bibpunct{(}{)}{;}{a}{}{,}

\usepackage{ulem}
\usepackage{lscape}
\usepackage{graphicx}
\usepackage[varg]{txfonts}
\def\d{$^\circ$}
\def\m{$^\prime$}

\def\cm3{cm$^{-3}$}

\def\2{$^{12}$CO}
\def\3{$^{13}$CO}
\def\8{C$^{18}$O}

\def\msol{M$_\odot$}

\def\cm2{cm$^{-2}$}

\begin{document}

\title{Radio study of HESS J1857$+$026. Gamma-rays from a superbubble?}
\author {A. Petriella \inst{1,2}
\and L. Duvidovich \inst{1,3}
\and E. Giacani \inst{1}
}

\institute{CONICET-Universidad de Buenos Aires, Instituto de Astronom\'ia y F\'isica del Espacio (IAFE), Buenos Aires, Argentina, 
\and Universidad de Buenos Aires, Ciclo B\'asico Com\'un, Buenos Aires, Argentina
   \email{apetriella@iafe.uba.ar}
\and Universidad de Buenos Aires, Facultad de Ciencias Exactas y Naturales, Buenos Aires, Argentina
}

\offprints{A. Petriella}

   \date{Received <date>; Accepted <date>}

\abstract{}{We provide new insights into the nature of HESS J1857$+$026, a very-high-energy $\gamma$-ray 
source whose complex morphology in the TeV band was attributed to the superposition of two 
distinct sources.}
{We performed radio continuum observations to look for the pulsar wind nebula and the supernova remnant associated with the pulsar 
PSR J1856$+$0245, which might be powering part of the $\gamma$-ray emission.  
We observed HESS J1857$+$026 with the Karl G. Jansky Very Large Array (VLA) 
at 1.5 GHz in the C configuration. In addition, using the same array configuration, we observed a region of $0.4^\circ \times 0.4^\circ$
towards PSR J1856$+$0245 at $6.0\,\mathrm{GHz}$. 
We obtained complementary data for the neutral hydrogen and molecular gas emission from public surveys in order to 
investigate the properties of the interstellar medium in the direction of HESS J1857$+$026.}
{The new observations at $1.5\,\rm{GHz}$ do not show evidence of emission above the noise level of $0.7\,\rm{mJy\,beam^{-1}}$ that
could be associated with either HESS J1857$+$026 or PSR J1856$+$0245. 
Also, in the new image at $6.0\,\rm{GHz}$ we do not detect radio emission from a pulsar wind nebula powered by PSR J1856$+$0245.
The neutral gas analysis shows the existence of a superbubble in the direction of the $\gamma$-ray source. 
We suggest that this structure is located at $\sim5.5\,\rm{kpc}$, compatible with the distance to the pulsar PSR J1856$+$0245.
}
{We conclude that TeV emission from HESS J1857$+$026 originates in a superbubble, arguing in favour of a single
$\gamma$-ray source rather than the superposition of two distinct sources. 
The pulsar PSR J1856$+$0245 could also be contributing as a source of $\gamma$-rays within the bubble.}
{}

\titlerunning{The nature of HESS J1857$+$026}
\authorrunning{A. Petriella et al.}


\maketitle

\section{Introduction}
\label{introd}

HESS J1857$+$026 was discovered by the H.E.S.S. observatory during a survey of the Galactic plane \citep{ahar08}. The source was clearly identified as extended. These latter authors obtained a centroid for the TeV emission at $(l,b)=(35.96,-0.06)$,
semi-major and -minor axes of about $0.11^\circ$ and $0.08^\circ$, respectively, and some emission projecting towards the north. At present, HESS J1857$+$026 has no clear counterpart in other spectral bands. It harbours the Vela-like radio pulsar PSR J1856$+$0245, 
located $\sim 8^{\prime}$ offset from the centroid of
the $\gamma$-ray emission. This latter is a $81\,\mathrm{ms}$ pulsar discovered by the Arecibo PALFA survey of the Galactic plane. The timing analysis in the radio domain shows that the pulsar has a spin-down age of $\sim 21\,\mathrm{kyr}$ and a spin-down
energy $\dot{E}\sim4.6 \times 10^{36}\,\mathrm{erg\,s^{-1}}$ \citep{hessels08}. 
A distance of $\sim 6.3\,\mathrm{kpc}$ was obtained using
the dispersion measure (DM) of $624\,\mathrm{cm^{-3}\,pc}$ and the electron density model YMW16 \citep{yao17} (we refer to the
ATNF Pulsar Catalog; \citealt{Manchester05}).
\citet{hessels08} suggest that HESS J1857$+$026 could be a pulsar wind nebula (PWN) powered by PSR J1856$+$0245,
based on the position and energetics of the  pulsar, which translates into a $\gamma$-ray efficiency $\epsilon_\gamma = L_\gamma / \dot{E,}$ 
similar to other TeV PWN/pulsar associations. 
This scenario is supported by the presence of the
ASCA X-ray source coincident with the radio pulsar. Regardless of the low number of counts, the authors derived a 
X-ray efficiency ($\epsilon_X = L_X / \dot{E}$) similar to other Vela-like pulsars powering 
synchrotron nebulae. \cite{nice13} performed a detailed study of the X-ray source using 
data from the {\it XMM-Newton} and {\it Chandra} telescopes. These authors found a point source coincident with PSR J1856$+$0245 and 
no evidence of extended emission related to a PWN. The spectrum is likely non-thermal and highly absorbed (indicating
a distant source) and would be produced by particles accelerated at the pulsar's magnetosphere.

In the GeV band, HESS J1857$+$026 was first detected by $Fermi$-LAT as a point source with a centroid coincident with the H.E.S.S. position 
\citep{rousseau12}. No pulsed emission from PSR J1856$+$0245 was reported in the $Fermi$ data.  
The broad band spectral analysis (from the X-ray to the TeV bands) reveals that the leptonic scenario is compatible with the 
pulsar's age and produces an acceptable fit assuming a low magnetic field ($\sim4\,\mathrm{\mu G}$).
The hadronic mechanism for $\gamma$-ray production also gives an adequate fit, but requires a very high injection energy.  
More recently, \citet{ackermann17} analysed 6 years of observations with $Fermi$-LAT above 10 GeV and 
found evidence of extended emission towards HESS J1857$+$026. Even if the extended nature of the source is undoubted, 
the large extent derived from the spatial analysis ($\sim 0.6^\circ$), which is indeed larger than the TeV extension, could be affected 
by two nearby hot spots likely unrelated with HESS J1857$+$026. 

The most detailed study of HESS J1857$+$026 in the TeV band was presented by \citet{aleksic14}, using data obtained with the
MAGIC telescope. These authors find that in the $0.3-1\,\mathrm{TeV}$ energy band  
the emission is well fitted with a 2D Gaussian function 
with a centroid coincident with that of H.E.S.S., while above $1\,\mathrm{TeV}$ the source is resolved into two distinct sources, 
labelled MAGIC J1857.2$+$0263 (hereafter MAG1) and MAGIC J1857.6$+$0297 (hereafter MAG2).
The authors propose a two-source model for HESS J1857$+$026, in which the $\gamma$-ray appearance is 
produced by the superposition in the line of sight of two unrelated sources located at two different distances.
   
On the one hand, the pulsar PSR J1856$+$0245 appears projected over MAG1, which presents an elongated shape.
The authors conclude that the pulsar is powering a PWN, which has the expected energy-dependent morphology in 
the very high-energy domain: the lower-energy electrons diffuse further from the pulsar before cooling through inverse Compton scattering, 
causing a smaller extension of the nebula for energies $>1\,\mathrm{TeV}$. 
Therefore, according to \citet{aleksic14}, MAG1 would be a TeV PWN located at $\sim9\,\mathrm{kpc}$, as they used the out-of-date NE2001 
electron density model of \citet{Cordes02} to derive the DM distance of the pulsar.
The elongated morphology of the PWN could be a consequence of the interaction with the reverse shock of the (so far undetected) 
supernova remnant (SNR) that was created together with the pulsar.

Regarding MAG2, which is resolved only above $1\,\mathrm{TeV}$ as a point source, its nature remains unclear. In the direction of
this source, \citet{aleksic14} found molecular clouds, HII regions, high-density gas, and H$_{2}$O maser emission; they also report on the detection of an incomplete shell of CO in the $54-56\,\mathrm{km\,s^{-1}}$ velocity range 
that partially overlaps MAG2.
The ultracompact HII region U36.40$+$0.02 (located at $\sim 3.7\,\mathrm{kpc}$) is probably related to this 
molecular structure. Based on the presence of HII regions, together with the detection of H$_{2}$O masers, 
massive stars and star forming activity are expected to be present in the field. These latter authors propose different scenarios 
to explain the nature of MAG2. 
TeV emission could arise from particles
accelerated in the colliding winds of massive stars or in the shocks produced by the outflows of protostars. 
Alternatively, the TeV radiation could be produced by electrons accelerated by an unidentified PWN evolving inside
the cavity blown by the wind of its progenitor star and/or by the HII region U36.40$+$0.02. 
As the peak of the TeV emission does not match the CO peaks, a leptonic mechanism for $\gamma$-ray production is favoured
with respect to a hadronic scenario. 

It is worth noting that the pulsar PSR J1857$+$0300 lies projected onto the TeV emission from HESS J1857$+$026, in the direction
of MAG2. Discovered by the Arecibo telescope \citep{lyne17}, it is an old ($\tau_c \sim 4.6 \times 10^6\,\mathrm{yr}$) 
and low-energy ($\dot{E} \sim 2.5 \times 10^{32}\,\mathrm{erg\,s^{-1}}$) neutron star. From the DM of $\sim 691\,\mathrm{pc\,cm^{-3}}$, 
a distance of $\sim 6.7\,\mathrm{kpc}$ is obtained using the YMW16 electron density model.

So far, the nature of HESS J1857$+$026 remains unclear and the two-source scenario of \citet{aleksic14} needs 
to be further investigated. For this purpose, we obtained high-quality radio observations using the 
Karl G. Jansky Very Large Array\footnote{The VLA is operated by the National Radio Astronomy Observatory (NRAO).} (VLA) 
covering the whole extension of the source.
We plan to detect the radio counterpart of the putative TeV PWN powered by PSR J1856$+$0245 and the shell of the associated SNR, 
allowing us to characterise the sources that may 
be accelerating particles and producing $\gamma$-rays. We also investigate the distribution of the neutral and molecular gas using 
data taken from public surveys to look for dense material that might correlate
with the $\gamma$-ray emission and that could be acting as an efficient proton target. Additionally, we plan to find evidence in the interstellar medium (ISM) of explosive and/or expanding events in the region. This will allow us to test the most recent models
of very-high-energy emission, which suggest that TeV-emitting particles are efficiently accelerated
by the combined action of stellar winds and SNRs in the interior of superbubbles (SBs).

\section{Data}

\subsection{New radio observations}

HESS J1857$+$026 was observed with the VLA in the C configuration (project ID 18B-001) using a mosaicing technique with 14 different pointings following a hexagonal pattern. The data were taken at L-band using the wide-band $1\,\mathrm{GHz}$ receiver system centred 
at $1.5\,\mathrm{GHz}$ which consists of 16 spectral windows, each with a bandwidth of $64\,\mathrm{MHz}$, spread into 64 channels. The source 3C 48 was used as a primary flux density and bandpass calibrator, while the phase was calibrated with J1822$+$0938. 
We also performed observations with the VLA in the C configuration towards a field of 0.4\d $\times$ 0.4\d~in the direction of 
PSR J1856$+$0245 using the mosaicing technique with 14 different pointings. These observations were planned to detect 
the putative PWN associated with the pulsar. In this case, we used the wide-band $4-8\,\mathrm{GHz}$ receiver system. 
We requested the 3-bit sampler with two sub-bands of $2\,\mathrm{GHz}$, each comprising 16 spectral windows 
with a bandwidth of $128\,\mathrm{MHz}$ each, spread into 64 channels. The source 3C 48 was used as flux density and bandpass calibrator, 
and J1922$+$1530 as a phase calibrator. Table \ref{Radio_observaciones_1857} summarises the observations.

\begin{table}[ht]
\centering
  \caption{Radio continuum observations with the VLA.}
    \begin{tabular}{ccc}
    \hline \hline
          & HESS J1857$+$026 & PSR J1856$+$0245 \\
    \hline
    Date & 17 Nov 2018 & 11 Jan 2019 \\
     & 24 Nov 2018 & 23 Feb 2019 \\
    Array configuration & C & C \\
    Frequency band & L & C \\
    Bandwidth & $1\,\mathrm{GHz}$ & $4\,\mathrm{GHz}$ \\
    Spectral windows & 16 ($64\,\mathrm{MHz}$) & 16 ($128\,\mathrm{MHz}$) \\
    Integration time & $25\,\mathrm{min}$ & $30\,\mathrm{min}$ \\
    Mapped region & $1.4^{\circ} \times 1.4^{\circ} $ & $0.4^{\circ} \times 0.4^{\circ} $ \\    
    Angular resolution & 17\m\m.38 $\times$ 15\m\m.51 & 3\m\m.55 $\times$ 2\m\m.62 \\
    Sensitivity (rms) & $0.7\,\mathrm{mJy\,beam^{-1}}$ & $12\,\mathrm{\mu Jy\,beam^{-1}}$ \\
    \hline
    \end{tabular}
\tablefoot{The {\it Integration time} refers to the total time on-source for each pointing of the mosaic and {\it Mapped region} indicates the rough sizes of the final maps at L band (Fig. \ref{radio_fig_ver}) and C band (Fig. \ref{Radio_pulsar_magic}).}
    \label{Radio_observaciones_1857}
\end{table}

All data were processed with the VLA CASA Calibration Pipeline. We applied extra flagging before imaging to improve the
quality of the calibrated data.
The final images were obtained with the TCLEAN task of the CASA software using the multi-frequency deconvolution algorithm for both 
spectral bands.
At L band we found structures with different spatial scales, so we used a multiple scale algorithm in the deconvolution procedure. 
The final image at L band has a resolution of $17^{\prime\prime}.38 \times 15^{\prime\prime}.51$, an angle of 6\d.87, and an 
effective noise of $0.7\,\mathrm{mJy\,beam^{-1}}$. 
The final image at C band has a resolution of $3^{\prime\prime}.55 \times 2^{\prime\prime}.62$, 
an angle of $-50^{\circ}.8$, and an effective noise of $12\,\mathrm{\mu Jy\,beam^{-1}}$.

\subsection{The surrounding medium}

The neutral hydrogen (HI) data were extracted from 
the Very Large Array Galactic Plane Survey (VGPS, \citealt{stil06}), 
which maps the $21\,\mathrm{cm}$ line emission with angular and spectral resolutions
of 1\m~and $1.3\,\mathrm{km\,s^{-1}}$, respectively.
We also studied the distribution of the molecular gas using the emission of the $^{13}$CO 
extracted from the Galactic Ring Survey (GRS, \citealt{jack06}). 
The survey maps the Galactic ring in the $^{13}\mathrm{CO\,J}$=$1-0$ line with angular and spectral resolutions 
of 46\m\m~and $0.2\,\mathrm{km\,s^{-1}}$, respectively.
Besides, we used the $^{12}\mathrm{CO\,J}$=$1-0$ emission from the FOREST Unbiased Galactic Plane Imaging (FUGIN) 
survey\footnote{Retrieved from the JVO portal (http://jvo.nao.ac.jp/portal/) operated by ADC/NAOJ.} \citep{umemoto17}. 
The angular and velocity resolutions are 20$^{\prime\prime}$ and $0.65\,\mathrm{km\,s^{-1}}$, respectively. 

\section{Results}

\subsection{The new VLA images} 

\begin{figure*}[ht]
\centering
\includegraphics[width=0.8\textwidth]{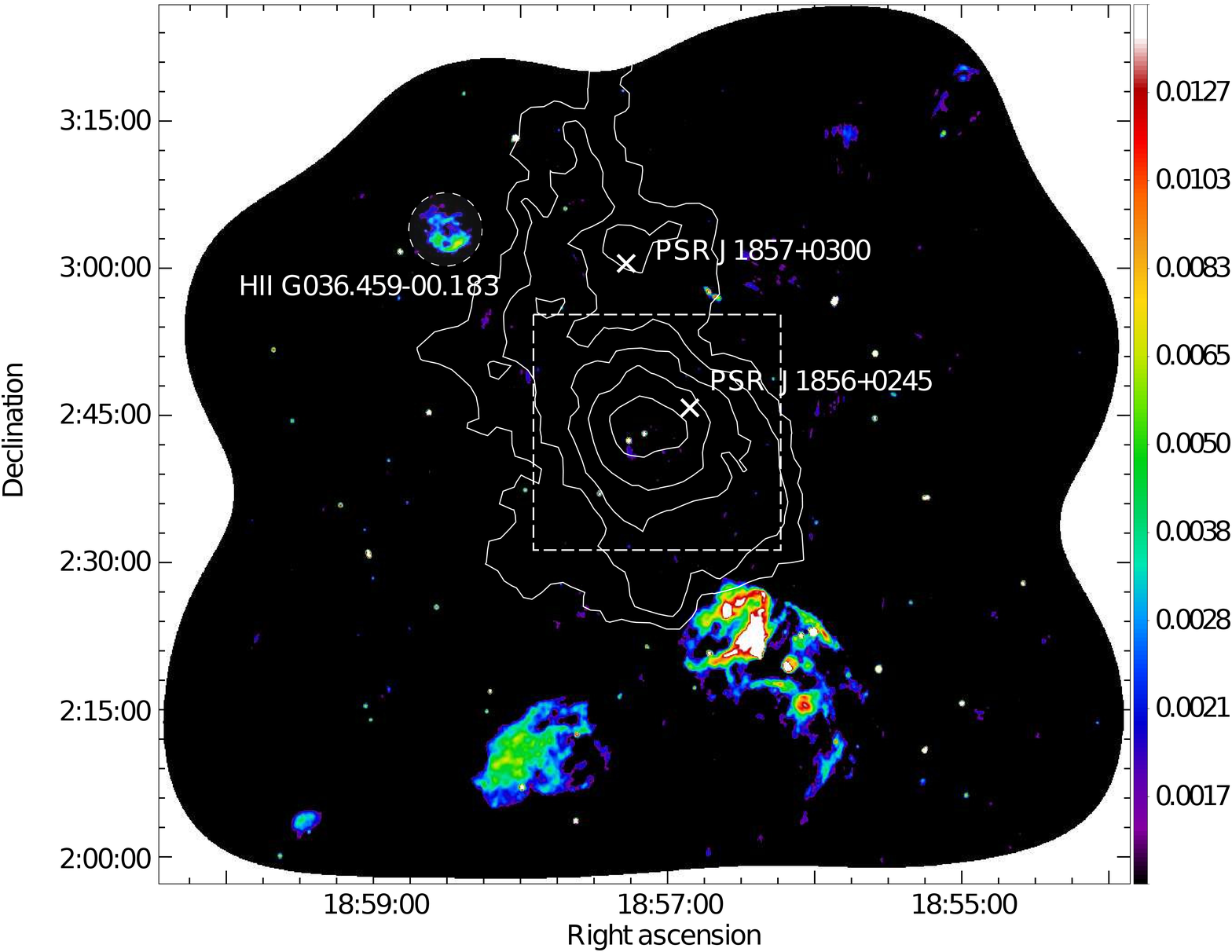}
\caption{Radio continuum emission at $1.5\,\mathrm{GHz}$. 
The white crosses indicate the positions of the pulsars PSR J1856$+$0245 and PSR J1857$+$0300. 
The white square shows the region mapped in the C band with the VLA and the contours represent the TeV emission below $1\,\mathrm{TeV}$ 
extracted from \citet{HESS18imagenes} with values of 1.5, 2.0, 2.7, 3.5, and 4.5 in units of $10^{-13}\,\mathrm{ph\,cm^{-2}\,s^{-1}}$.
The angular resolution is 17\m\m.38 $\times$ 15.51\m\m, 
$\rm{PA}=6^{\circ}.87$ and the noise is $0.7\,\mathrm{mJy\,beam^{-1}}$. The color scale is expressed in $\mathrm{Jy\,beam^{-1}}$.}
\label{radio_fig_ver}
\end{figure*}

The new radio image at $1.5\,\mathrm{GHz}$ is shown in false colour in Fig. \ref{radio_fig_ver} with contours of the $\gamma$-ray emission
for energies $<1\,\mathrm{TeV}$. 
The positions of the pulsars PSR J1856$+$0245 and PSR J1857$+$0300 are indicated with white crosses.
There is no evidence, down to the noise level of the image, 
of enhanced emission in the direction of PSR J1856$+$0245 that
could represent a probable PWN, or of the surrounding radio shell associated with the host SNR. 

\begin{figure*}[ht!]
\centering
\includegraphics[width=0.8\textwidth]{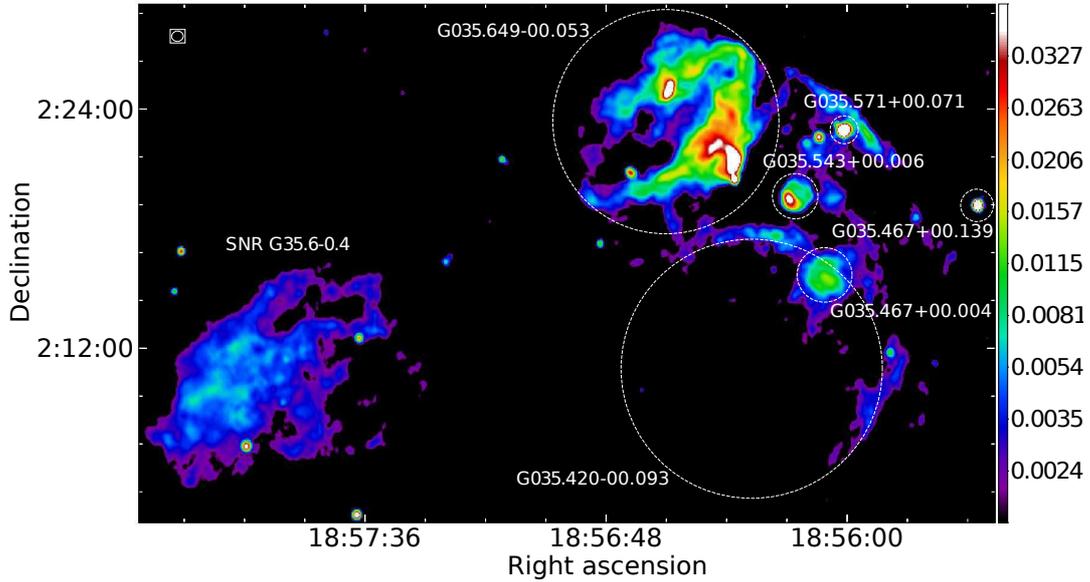}
\caption{Magnification of the southern region of the image obtained at 1.5 GHz. 
The circles represent the positions of the HII regions
taken from the WISE Catalog of Galactic HII Region.
The synthesized beam is represented in the upper left corner. The color scale is expressed in $\mathrm{Jy\,beam^{-1}}$.}
\label{Radio_yhii}
\end{figure*}

As can be seen from Fig. \ref{radio_fig_ver}, several extended sources are detected in the field. 
To the north, we detect the HII region G036.459$-$00.183, which has a systemic velocity of $\sim73\,\mathrm{km\,s^{-1}}$ and
a kinematic distance of $\sim8.9\,\mathrm{kpc}$ (refer to the WISE Catalog of Galactic 
HII Regions\footnote{Available at http://astro.phys.wvu.edu/wise/.} and references therein). 
To the south, we identify the SNR G35.6$-$0.4 and a complex of HII regions. 
This complex has a systemic velocity of $\sim52\,\mathrm{km\,s^{-1}}$ and a kinematic 
distance of $\sim10.4\,\mathrm{kpc}$ (see the aforementioned catalogue of HII regions).
In Fig. \ref{Radio_yhii} we display a zoomed view of the area.


\begin{figure*}[ht!]
\centering
\includegraphics[width=0.8\textwidth]{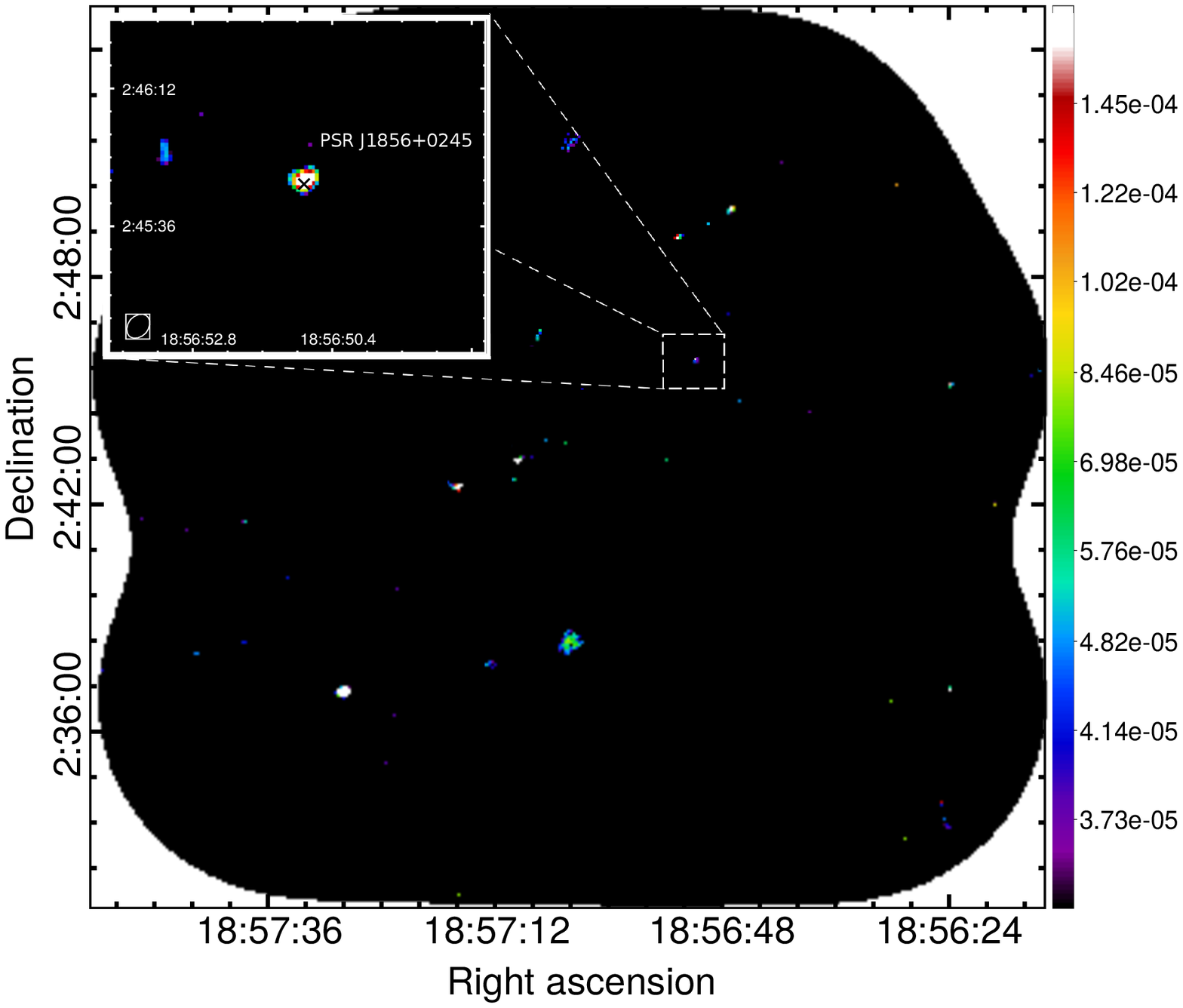}
\caption{Radio continuum emission at $6.0\,\mathrm{GHz}$. 
The resolution is 3\m\m.55 $\times$ 2\m\m.62 while the 
effective noise level is $12\,\mathrm{\mu Jy\,beam^{-1}}$. The color scale is expressed in $\mathrm{Jy\,beam^{-1}}$.}
\label{Radio_pulsar_magic}
\end{figure*}

This is another case in which the new dedicated radio observations carried out with higher sensitivity than 
current surveys do not show any indication of diffuse emission around the pulsar.
In Section \ref{secc_disc}, we discuss the possible reasons for this. In principle, we rule out that it is a consequence of the limitations of the interferometer when observing extended sources. On the one hand, the largest angular scale of the VLA in the C configuration at L band is $\sim 8^{\prime}$, similar to the extent of the source MAG1 ($10^{\prime}\times 4^{\prime}$), 
and so we would expect no significant losses of flux from a putative PWN of a similar size. 
On the other hand, considering the large extent of HESS J1857$+$026 with respect to the primary beam 
($\theta_{PB}\sim 17^{\prime}$ FWHM at L band), the mosaicing technique guarantees a constant sensitivity over the field, avoiding the attenuation of the primary beam.

The radio continuum emission at $6.0\,\mathrm{GHz}$ is shown in Fig. \ref{Radio_pulsar_magic}. 
The new high-sensitivity radio map does not reveal any extended emission around PSR J1856$+$0245. 
We detect a point source at the position of the pulsar,  
with a flux density $S_{6.0\,\mathrm{GHz}}\sim0.23\,\mathrm{mJy}$. 
A zoomed-in view of this region is shown in the upper left corner of Fig. \ref{Radio_pulsar_magic}.
The flux of the pulsar at $1.4\,\mathrm{GHz}$ was measured by \citet{nice13}, who obtained $S_{1.4\,\mathrm{GHz}}=0.58\pm0.02\,\mathrm{mJy}$. 
We note that the flux is below the sensitivity of our $1.5\,\mathrm{GHz}$ observation (Table \ref{Radio_observaciones_1857}), 
explaining the non-detection of the pulsar in the new VLA image of Fig. \ref{radio_fig_ver}. 
 \citet{nice13} also derived a spectral index $\alpha = 0.4\pm0.2$ ($S_{\nu}\propto\nu^{-\alpha}$) by fitting the emission of their 
four $100\,\mathrm{MHz}$ sub-bands.     
We estimate the spectral index by combining the flux at $1.4\,\mathrm{GHz}$ from \citet{nice13} and the
flux at $6.0\,\mathrm{GHz}$ from our observations. We obtain $\alpha = 0.64\pm0.06$ for PSR J1856$+$0245.

\subsection{Neutral gas distribution}
\label{HI}

We investigated the neutral gas distribution towards HESS J1857$+$026, looking for: (i) high-density gas 
that correlates with the TeV emission and could be acting 
as an efficient proton `target' for hadronic production of $\gamma$-rays, and (ii) the vestiges in the ISM of the action of expansive or explosive events, such as
the explosion of the SN that could have created the pulsars PSR J1856$+$024 and PSR J1857$+$0300.

\begin{figure*}[ht]
\centering
\includegraphics[width=0.7\textwidth]{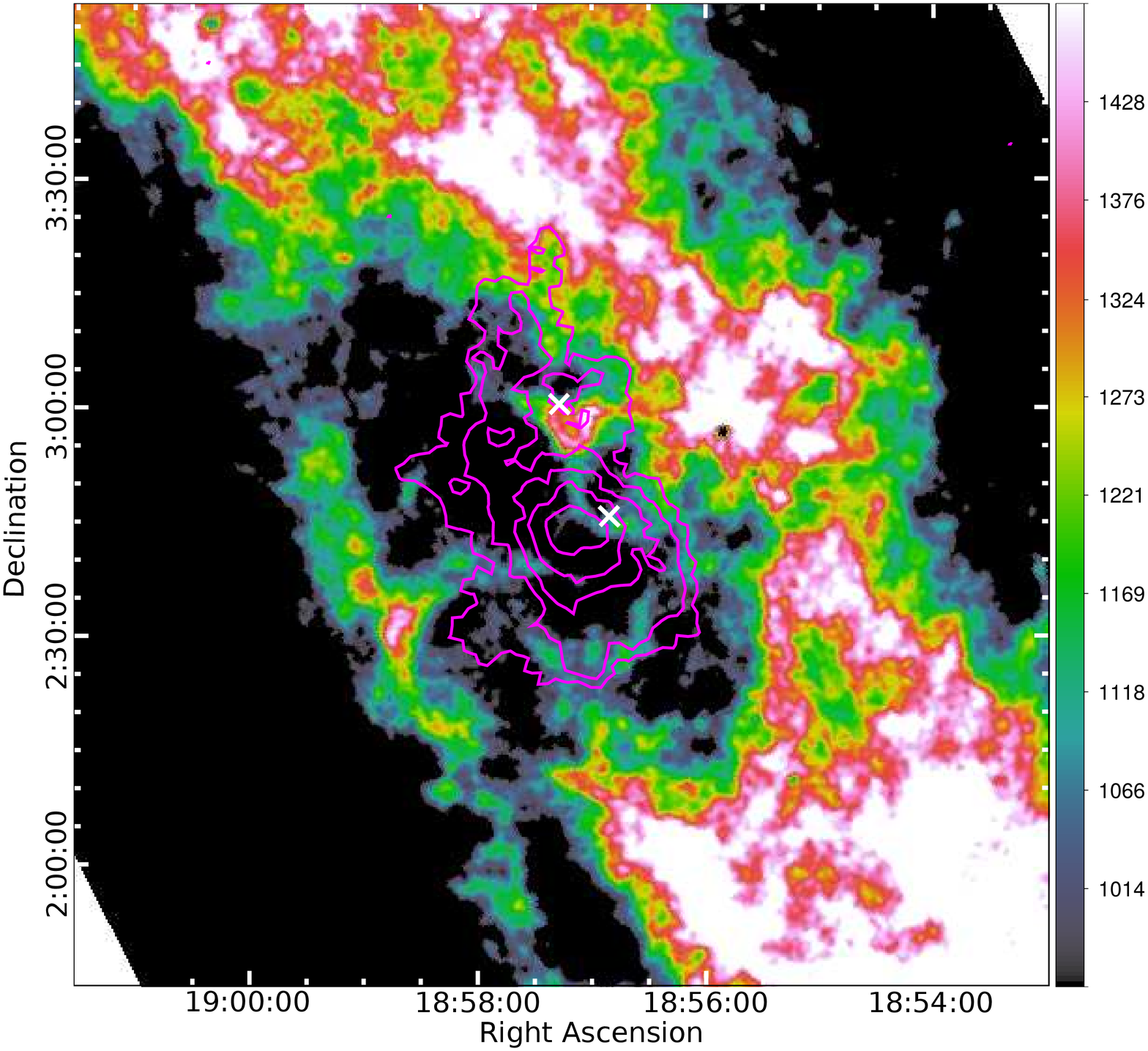}
\caption{HI distribution towards HESS J1856$+$026, 
integrated in the velocity range between 81 and $102\,\mathrm{km\,s^{-1}}$. 
The contours correspond to the TeV emission and the crosses mark the position of the pulsars PSR J1856$+$0245 and PSR J1857$+$0300.
The color scale is expressed in arbitrary units.}
\label{HI_HESS_PULSAR}
\end{figure*}

After inspecting the complete HI cube, we found a cavity-like structure surrounded by an incomplete 
shell of neutral gas in the velocity range between 81 and $100\,\mathrm{km\,s^{-1}}$. 
Figure \ref{HI_HESS_PULSAR} shows the integration of the HI emission in this interval, together with contours of the TeV
emission from HESS J1857$+$026, and the positions of the pulsars PSR J1856$+$0245 and PSR J1857$+$0300. 
Interestingly, the size and morphology of the cavity approximately match those of the TeV emission. Both of them display an
elliptical shape elongated in the direction parallel to the Galactic plane.  
Adopting a systemic velocity of $91\,\mathrm{km\,s^{-1}}$ for the HI structure, and using the rotation model of \citet{persic1996} with the parameters of \cite{reid2014}, we obtained near and far distances of 5.5 and $\sim8.3\,\mathrm{kpc}$, respectively. 
Taking into consideration the distance uncertainties of the kinematical method, the near distance is compatible
with the DM distances of the pulsars PSR J1856$+$0245 ($\sim6.3\,\mathrm{kpc}$) and PSR J1857$+$0300 ($\sim6.7\,\mathrm{kpc}$).

Without any possibility of discriminating between the two kinematic distances, we assume as a working hypothesis
that the HI cavity and shell were swept by the expansion of the SNR created together with the pulsar and/or by the action of the winds
of its progenitor star. Then, following the association between the pulsars and the HI structure, we adopt  $5.5\,\mathrm{kpc}$ as the 
most probable distance to the cavity of neutral gas.

To characterise this HI structure, we estimate the neutral gas mass integrating the HI column density between 
81 and $100\,\mathrm{km\,s^{-1}}$. Assuming that the emission is optically thin, 
the HI column density is given by $N_{HI} = 1.8 \times 10^{18} \int(\Delta T dv)\,\mathrm{cm^{-2}}$, where $\Delta T$ is the mean 
brightness temperature of the integration region and $dv$ the velocity interval of integration. 
The mass is therefore 
\begin{equation}
M = 1.3 \times 10^{-3}D_{kpc}^{2}\Omega\Delta T dv~\rm{M}_\odot
\label{equ1}
,\end{equation} 
where $D_{kpc}$ is the distance in kiloparsecs and $\Omega$ is the solid angle of the integration region.

We approximate the shape of the shell with an elliptical ring with 
inner and outer boundaries of 35.5\m~$\times$~18\m~and 
55.8\m~$\times$~35\m, respectively. We note that the inner boundary defines the outer limit of the HI cavity.
Then, the mean HI brightness temperature of the swept shell integrated between 81 and $100\,\rm{km\,s^{-1}}$ 
is $T_{s}=1265\,\mathrm{K}$. 
The emission from the shell must be corrected 
with the background emission, that is, the emission of the ISM prior to the formation of the shell.
It follows that $\Delta T_{s} = T_{s} - T_{bg}$, where $T_{bg}$ is the background temperature.
A reasonable assumption to obtain $T_{bg}$ is to consider it equal to the temperature of the unperturbed ISM just 
outside the shell. From Fig. \ref{HI_HESS_PULSAR} we note that the HI distribution around the shell is very inhomogeneous 
and it is difficult to choose an appropriate background region. Therefore, we followed the method of \cite{Suad2019AA} to derive $T_{bg}$, 
which assumes that all the mass evacuated from the cavity ($M_{cav}$) is accumulated in the shell ($M_{s}$) and that 
$\Delta T_{cav}=T_{bg}-T_{cav}$. For the HI cavity, we derived a mean HI brightness temperature $T_{cav}=970\,\mathrm{K}$.
Therefore, using Eq. \ref{equ1}, the condition $M_{cav}= M_{s}$ gives 
$(T_{s} - T_{bg})\Omega_{shell} = (T_{bg} -T_{cav})\Omega_{cav}$. The latter can be solved to obtain $T_{bg}$.
Taking $T_{s}=1265\,\mathrm{K}$, $T_{cav} = 970\,\mathrm{K}$, $\Omega_{s} = 3370\,\mathrm{arcmin^{2}}$, and 
$\Omega_{cav} = 2010\,\mathrm{arcmin^{2}}$, we obtain $T_{bg}=1155\,\mathrm{K}$, and therefore $\Delta T_{s}=110\,\mathrm{K}$.

Using the mean HI brightness temperature of the shell, the column density of the neutral hydrogen is 
$N_{HI}\sim 1.6 \times 10^{20}\,\mathrm{cm^{-2}}$ and the mass of the swept shell is $M_{s}\sim 1.6 \times 10^4\,\mathrm{M_\odot}$.
To estimate the original ambient density $n_0$, we performed the integration of Eq. \ref{equ1} over the cavity area 
using the background temperature, which is the temperature of the unperturbed ISM prior to the expansion of the HI shell. 
The obtained mass is distributed in the volume of the cavity, which is
approximated by a sphere of radius $R_{eff}=\sqrt{ab}$, where $a$ and $b$ are the mean major and minor semi-axis
of the shell.
Then, for $R_{eff} = 56\,\mathrm{pc}$, we obtain an original ambient density $n_{0}\sim7.5\,\mathrm{cm^{-3}}$.
The kinetic energy is derived from $E_k = 1/2M_{s}v_{exp}^{2}$ and the dynamical age 
from $t_{dyn}= 0.6R_{eff}/v_{exp}$, where $v_{exp}$ is the expansion velocity of the shell. 
As we show later, the cavity was probably created by stellar winds, so a factor 0.6 
is included according to evolutionary models of interstellar bubbles \citep{weaver77}.  
We take $v_{exp}=10\,\mathrm{km\,s^{-1}}$, which is half the velocity range where the HI shell is detected. 
This expansion velocity is a lower limit, because
confusion from unrelated foreground or background emission
hampers the detection of the caps of the expanding shell.
We obtain $E_{k} \sim 1.6 \times 10^{49}$ erg and $t_{dyn} \sim 3.3 \times 10^{6}\,\mathrm{yr}$.
Generally, this method produces large uncertainties of the parameters (up to 40$\%$) 
due to the error in the distance, the choice of the background level, and the integration limits. 
Despite the errors involved, the age of the HI shell is two orders of magnitude greater than the age of 
PSR J1856$+$0245. The parameters that characterise the shell are summarised in Table \ref{parameters_1857}.

\begin{table}[ht]
\centering
  \caption{Observed and estimated parameters of the HI superbubble.}
    \begin{tabular}{cc}
    \hline \hline
    Shell Centre & RA: 18$^{\rm{h}}$ 57$^{\rm{m}}$ 22$^{\rm{s}}$ \\
     & DEC. = 2\d 43\m 13\m\m.6 \\
    Size (semi-axis) & 46\m $\times$ 27\m \\
    Central velocity & $91\,\mathrm{km\,s^{-1}}$ \\
    Expansion velocity ($v_{exp}$) & $10\,\mathrm{km\,s^{-1}}$ \\
    HI column density ($N_{\rm{HI}}$) & $1.6 \times 10^{20}\,\mathrm{cm^{-2}}$ \\
    Swept mass ($M_{s}$) & $1.6 \times 10^{4}\,\mathrm{M_\odot}$ \\
    Original density ($n_0$) & $7.5\,\mathrm{cm^{-3}}$ \\
    Kinetic energy ($E_k$) & $1.6 \times 10^{49}\,\mathrm{erg}$ \\
    Dynamical age ($t_{dyn}$) & $3.3\,\mathrm{Myr}$ \\
    \hline
    \end{tabular}
    \label{parameters_1857}
\end{table}

The derived morphological and physical characteristics of the HI shell are similar to those found in 
Galactic SBs \citep{Suad2014AA,Suad2019AA}. 
The origin of these structures is generally attributed to the action of the stellar winds of massive stars and their 
subsequent explosion as SNe. As neither associations of OB stars nor other types of early stars were found in the literature within the limits of the cavity, we analyse the types of stars that could have formed the SB. We compare the kinetic energy $E_k$ of the SB with the mechanical energy $E_w$ injected by the stellar wind into the ISM during the lifetime of the bubble.
The latter is estimated from the relation $E_w=\epsilon_w L_w t_{dyn}$, where $\epsilon_w$ is the fraction of the stellar wind that is converted into mechanical energy 
and $L_w$ is the mechanical luminosity of the wind, defined
as $L_w = 1/2 \dot{M} v^2_w$, where $\dot{M}$ is the mass-loss rate of the star and $v_w$ is the terminal velocity of the wind.
Taking typical values of the stellar wind of an O-type star ($\dot{M} = 2 \times 10^{-6}\,\mathrm{M_\odot\,yr^{-1}}$, 
$v_w= 2000\,\mathrm{km\,s^{-1}}$) \citep{prinja90,mokiem07}, we obtain $L_{O-star} \sim 2.5 \times 10^{36}\,\mathrm{erg\,s^{-1}}$.
Considering a canonical $\epsilon_w = 0.2$ \citep{weaver77}, the total energy released during the lifetime of the SB 
($t_{dyn} = 3.3\,\mathrm{Myr}$) that is converted into mechanical energy is $E_w\sim 5 \times 10^{49}\,\mathrm{erg}$, which exceeds 
the kinetic energy of the shell.
This result supports a stellar wind origin for the SB, showing that a single massive star is energetic enough to evacuate the 
HI cavity and accumulate the material in the shell.  

\subsection{Molecular gas distribution}
\label{CO}

\begin{figure*}[ht]
\centering
\includegraphics[width=0.7\textwidth]{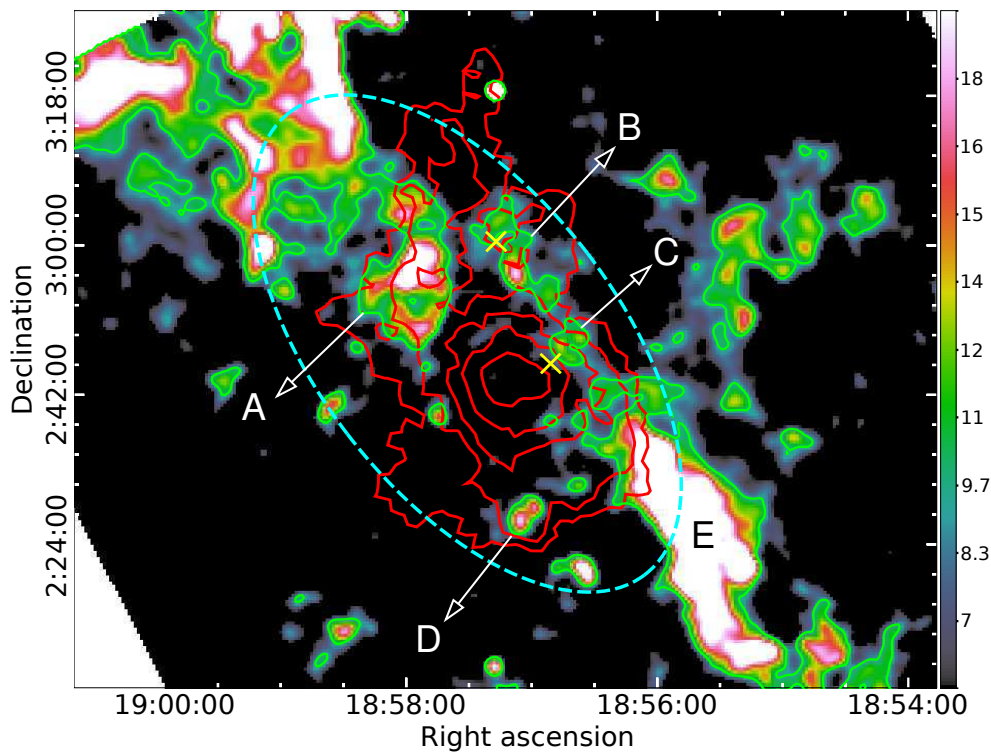}
\caption{$^{13}$CO emission integrated between 78 and $90\,\mathrm{km\,s^{-1}}$ with TeV emission
from HESS J1857$+$026 (red contours). The color scale is expressed in arbitrary units. 
The cyan ellipse indicates the inner boundary of the HI cavity,
and the yellow cross the positions of the pulsars PSR J1856$+$0245 and PSR J1857$+$0300.
The color scale is expressed in arbitrary units.}
\label{fig_CO}
\end{figure*}

We inspected the distribution of the $^{13}$CO J=1--0 in a wide area surrounding the VHE source and in the velocity interval
where the HI shell spans. We found several molecular condensations
between $78$ and $90\,\mathrm{km\,s^{-1}}$, which seem to form an incomplete shell around the brightest $\gamma$-ray emission
(see Fig. \ref{fig_CO}). 
We have labelled five molecular components: cloud A is located to the east of the TeV peak; 
clouds B and C, to the north; cloud D is a compact feature to the southwest; and cloud E is a large feature partially overlapping the 
TeV emission.

To derive the physical properties of the molecular clouds, we consider that the gas is at local thermodynamic equilibrium (LTE).
The $^{13}$CO J=1--0 line is assumed to be optically thin, and we use it to derive the H$_2$ column density $N(H_2)$. The 
excitation temperature $T_{ex}$ was obtained from the brightness temperature $T_B^{12CO}$ of the $^{12}$CO J=1--0 (taken from the FUGIN survey), which is assumed to be optically thick. It follows that $T_{ex}=5.5 [ln(1+5.5(T_B^{12CO}+0.82))]^{-1}$, where all the 
temperatures are expressed in K. We obtained a mean $T_{B}^{12CO}\sim4\,\mathrm{K}$ over clouds A, B, C, and D, which 
yields $T_{ex} \sim 7.7\,\mathrm{K}$. For cloud E, we obtain $T_{B}^{12CO}\sim 6.5\,\mathrm{K}$ and $T_{ex} \sim 9.8\,\mathrm{K}$. 
The $^{13}$CO column densities and H$_2$ masses were derived using equations 1, 2, and 3 of \citet{petriella12}.
We took a background temperature $T_b = 2.7\,\mathrm{K}$, a H$_2$ to $^{13}$CO relative abundance of $5 \times 10^5$ \citep{ortega2017}, 
and a helium abundance of 25\%. The projected shapes of the clouds were approximated by ellipses. To estimate the 
number densities, we approximated the cloud volume by ellipsoids with a third axis equal to the mean of the ellipse
semi-axis. We use a distance of $5.5\,\mathrm{kpc}$, coincident with the distance of the HI cavity. 
The results are listed in Table \ref{tablaCO}.  
 
\begin{table*}[ht]
\centering
  \caption{Observed and derived parameters of the molecular gas calculated for a distance of $5.5\,\mathrm{kpc}$.  }
    \begin{tabular}{cccccc}
    \hline \hline
     Cloud & A & B & C & D & E \\
    \hline
    Size (semi-axis) & 9.9\m$\times$5.6\m & 4.1\m$\times$1.8\m & 2.7\m$\times$2.7\m & 2.9\m$\times$2.0\m & 20.2\m$\times$5.8\m \\
    $N(H_2)$ ($\times 10^{21}\,\mathrm{cm^{-2}}$) & 2.9 & 2.4 & 2.3 & 2.6 & 4.3 \\
    Mass ($\times 10^4$ \msol) & 2.9 & 0.3 & 0.3 & 0.3 & 9.1 \\
    Density (cm$^{-3}$)  & 58 & 125 & 131 & 163 & 51\\
    \hline
    \end{tabular}
    \label{tablaCO}
\end{table*}

\section{Discussion}
\label{secc_disc}

The source first identified as a possible origin of $\gamma$-ray emission from HESS J1857$+$026 is the pulsar 
PSR J1856$+$0245. However, the complex morphology in the TeV band suggests that it 
could be formed by two different astrophysical objects \citep{ahar08}. 
On the one hand, \citet{aleksic14} suggest that MAG1 could be a PWN powered by PSR J1856$+$0245 and
located at $\sim9\,\mathrm{kpc}$.
On the other hand, MAG2, which is only resolved above $1\,\mathrm{TeV}$, 
appears as an extension of the most intense TeV emission from MAG1 and 
has no clear counterpart in other wavelengths, but it may be associated with
a cavity of CO and a UC HII region located at $\sim3.7\,\mathrm{kpc}$. 
This two-source scenario resembles that of VER J1907$+$062, whose complex morphology in the TeV band was explained by the 
superposition of two unrelated $\gamma$-ray sources located at two different distances \citep{crestan21,duvi20}.

Our analysis of the distribution and kinematics of the atomic gas revealed the existence of a SB towards HESS J1857$+$026
in the velocity range between 81 and $100\,\mathrm{km\,s^{-1}}$. 
Superbubbles are formed by the injection of a large sum of energy from stellar winds and SN explosions and were proposed as 
alternative sources of Galactic cosmic rays, in contrast to isolated SNRs (see e.g. \citealt{Parizot2004AA, Binns2007SSR}).
Indeed, the collective action of massive stars and SNRs within SBs could energise protons up to PeV energies, making
them the most promising pevatron candidates \citep{Ahar19}. 
In the context of the discovery of a SB coincident with HESS J1857$+$026, we analysed the possible contributions 
of all astrophysical objects to the $\gamma$-ray emission.

We obtained a kinematic distance of $\sim5.5\,\mathrm{kpc}$ for the SB, which is compatible, within the errors, 
with the DM distance of PSR J1856$+$0245 ($\sim6.3\,\mathrm{kpc}$).
The TeV emission appears completely immersed within the HI cavity and both of them present
elongated shapes in the direction parallel to the Galactic plane. 
The spatial and morphological match between the SB  
and the entire TeV emission from HESS J1857$+$026 suggests an association between them, arguing in favour 
of a single $\gamma$-ray source confined inside the HI cavity and located at a distance of $\sim 5.5\,\mathrm{kpc}$.
Even if there is no unequivocal evidence of the physical association between HESS J1857$+$026 and the neutral gas structure, we 
hypothesise that the SB may have been created by the stellar winds of massive stars located near its centre, with the
possible contribution of the SNRs formed after the explosion of at least one of these stars.
Such an SN event is suggested by the presence of the pulsar PSR J1856$+$0245.

In our scenario, PSR J1856$+$0245 is located inside the SB and part of the $\gamma$-ray emission from HESS
J1857$+$026 may correspond to a TeV PWN powered by the pulsar. 
The non-detection of a radio PWN resembles other energetic pulsars powering PWNe detected in the 
high- and very high-energy domains but not observed in the radio band. 
Examples of such pulsars are PSR J1907$+$0602 \citep{duvi20}, PSR J1826$-$1256 \citep{duvi19}, 
and PSR J1809$-$1917 \citep{castelletti2016};
all of them, like PSR J1856$+$0245, have magnetic fields of a few $10^{12}\,\mathrm{G}$, as commonly found in young pulsars.

Despite the non-detection of a radio PWN around PSR J1856$+$0245, we use the radio observations to 
put some constraints on its characteristics.
A rough estimate of the flux density of the PWN at $1.4\,\mathrm{GHz}$ is calculated from
$F_{1.4}[\textrm{mJy}]=2.15\times 10^8 \epsilon_R \dot{E}_{37} / D_{kpc}^2$
\citep{Gaensler00}, where $\epsilon_R$ is the efficiency of conversion of the pulsar power into radio emission, 
$\dot{E}_{37}$ is the spin-down power of the
pulsar in units of $10^{37}\,\mathrm{erg\,s^{-1}}$, and $D_{kpc}$ the distance in kiloparsec 
(taken equal to the SB distance of $5.5\,\mathrm{kpc}$). 
We obtain $F_{1.4}\sim330\,\mathrm{mJy}$. We assume that the radius of the radio PWN is $\sim 0.1^{\circ}$ ($=10\,\mathrm{pc}$ for a 
distance of $5.5\,\mathrm{kpc}$), similar to the spatial extension of the TeV source MAG1. 
Taking a typical $\epsilon_R = 10^{-4}$ \citep{gaensler2006book},
the expected surface brightness at $1.5\,\mathrm{GHz}$ is $S_{1.5}=F_{1.5}/A\sim0.6\,\mathrm{mJy\,beam^{-1}}$.
The latter was obtained after expressing the area $A$ of the PWN in units of the beam at $1.5\,\mathrm{GHz}$ 
(Table \ref{Radio_observaciones_1857}) and taking $F_{1.5}\sim F_{1.4}$.
Under all the assumptions made, the radio PWN remains undetectable as its surface brightness falls below the sensitivity of the 
radio image ($rms\sim0.7\,\mathrm{mJy\,beam^{-1}}$). For an even larger PWN, the radio emission is expected to be dimmer. 
On contrary, if the radio PWN is more compact, we expect a brighter radio nebula. 
For example, taking a radius of $\sim 0.05^{\circ}$ ($=5\,\mathrm{pc}$), 
we get $S_{1.5}\sim2.5\,\mathrm{mJy\,beam^{-1}}$, which is three times the sensitivity of our data. 
In this case, the non-detection could be explained by the shift of the particle injection of the pulsar to higher energies,
which translates into $\epsilon_R\sim 10^{-5}$ \citep{Gaensler00}. The non-detection of the radio PWN 
around PSR J1856$+$0245 could therefore be a consequence
of its large extent and/or an inefficient conversion of the pulsar power into radio emission.

The pulsar PSR J1857$+$0300, with a DM distance of $\sim6.7\,\mathrm{kpc}$, also appears projected over the SB. 
Its age ($\sim4.6\,\mathrm{Myr}$) is similar to the dynamical age of the SB ($\sim3.3\,\mathrm{Myr}$).
We cannot therefore discard that the SN explosion that gave birth to PSR J1857$+$0300 has occurred within the SB at its early formation.
Due to the kick velocity acquired after the SN event (which may be some hundreds of $\mathrm{km\,s^{-1}}$), the pulsar might have 
overcome the SB and now is observed distant from it. 
The contribution of PSR J1857$+$0300 to the $\gamma$-ray emission from HESS J1857$+$026 is certainly negligible as its
spin-down power ($\sim2.5\times10^{32}\,\mathrm{erg\,s^{-1}}$) is three orders of magnitude less than the TeV luminosity.    

We have shown that there is molecular material projected into the HI cavity. This material seems to form an incomplete 
shell around the peak of the TeV emission. We can speculate two origins for this material. 
Either it could be preexisting molecular material that was reached by the expanding SB and was not swept away or evaporated by it, 
or it could be material from the parent molecular cloud that was swept away by the stellar wind of the central stars 
and now is detected as an expanding shell of molecular material. 
As \citet{Parizot2004AA} point out, the interaction of the stellar winds inside SBs with high-density clumps can 
amplify turbulence and MHD waves.
In addition, dense gas could be acting as a target for hadronic interactions and subsequent production of $\gamma$-rays.
The explosion that gave birth to PSR J1856$+$0245 has left a SNR which remains undetectable in the radio band, despite the
improved quality of the new VLA observations. 
This remnant is expected to be evolving inside the SB and could be an efficient accelerator of cosmic rays. 
In a hadronic scenario, the integral flux of $\gamma$-ray photons with energies greater than $E$ is given by
$F_{\gamma}(>E) = 10^{-10} f_{\Gamma} E_{TeV}^{-\Gamma+1} A$ \citep{torres03}, 
where the factor $f_{\Gamma}$ depends on the spectral index $\Gamma$, 
$E_{\rm{TeV}} = E / \rm{TeV}$, and $A= \theta E_{51} D_{kpc}^2 n_0\,\mathrm{ph\,cm^{-2}\,s^{-1}}$, where $\theta$ 
is the efficiency of conversion of the SN energy into cosmic ray energy, $E_{51}$ is the mechanical energy of 
the SN explosion in units of $10^{51}\,\mathrm{erg}$, $D_{kpc}$ is the distance in kiloparsecs, and $n_0$ the ambient density of 
the ISM in cm$^{-3}$.
Spectral data from HESS H1857$+$026 were taken from \citet{HGPS}: $\Gamma \sim 2.57$ (for which, we took $f_{\Gamma} = 0.19$) and
$F_{\gamma}(> 1\,\rm{TeV}) \sim 4 \times 10^{-12}\,\mathrm{ph\,cm^{-2}\,s^{-1}}$.
For a canonical SN explosion ($E_{51}=1$) at $5.5\,\mathrm{kpc}$ and $\theta = 0.3$, an ambient density $\sim22\,\mathrm{cm^{-3}}$ 
is necessary to produce the $\gamma$-ray flux from HESS J1857$+$026. Therefore, based on the density derived from the $^{13}$CO emission 
(Table \ref{tablaCO}), all the clouds are dense enough to act as a target for hadronic emission associated with the SNR.
If this mechanism contributes significantly to the TeV emission from HESS J1857$+$026, 
we would expect a coincidence between the distribution of the $\gamma$-ray emission and the molecular gas. 
Figure \ref{fig_CO} shows that the $^{13}$CO overlaps with the $\gamma$-ray source, but we do not detect an enhancement of the TeV
emission at the position of the molecular clouds. The hadronic mechanism associated with the putative SNR 
could therefore be contributing to the TeV emission from HESS J1857$+$026 but further evidence is necessary to prove this scenario.

Finally, the SB appears surrounded by the HII region G036.459$-$00.183 to the north and 
a complex of HII regions to the south (see Figs. \ref{radio_fig_ver} and \ref{Radio_yhii}), suggesting
that the expanding bubble may have triggered the formation of a second generation of stars.
There is evidence of star forming activity in the shells of SBs \citep{suad16,suad12}.
This is not straightforward for the SB associated with HESS J1857$+$026.
On the one hand, G036.459$-$00.183 is located at a distance of $\sim8.9\,\mathrm{kpc}$, 
which is not compatible with the distance of the SB.
On the other hand, the HII complex is probably located at $\sim10.4\,\mathrm{kpc}$ based on the distance estimate
of G035.649$-$00.053 obtained from the association with molecular gas at $\sim 52\,\mathrm{km\,s^{-1}}$ \citep{anderson09b}. 
However, we note that the HII region G035.467$+$00.139 is not related with the complex. There is associated molecular 
gas at a velocity of $\sim77.2\,\mathrm{km\,s^{-1}}$, which places this HII region at $\sim4.6\,\mathrm{kpc}$, compatible 
with the distance of the SB. 
A detailed study of the star forming activity is required to determine whether or not the expanding HI shell is triggering the formation
of new stars.
  
\section{Conclusions}
\label{secc_concl}

Following the discovery of a SB in the direction of HESS J1857$+$026, we analysed an alternative scenario
to the two-source model proposed by \citet{aleksic14} to explain the very-high-energy emission.  

The positional and morphological coincidence between the SB  
and the TeV emission from HESS J1857$+$026 suggests an association between them and 
argues in favour of a single $\gamma$-ray source rather than the separation into two different sources at different distances.
SBs are created by the collective effects of stellar winds and SNRs and are considered cosmic-ray factories and 
sources of $\gamma$-ray emission in the GeV and TeV bands.
We argue that the pulsar PSR J1857$+$0245 is associated with the SB and 
could also be contributing to the $\gamma$-ray emission. 
Indeed, this pulsar is energetic enough to power the entire TeV emission from HESS J1857$+$026. 
The SNR associated with this pulsar remains undetected, but due to its young age it is probably still expanding 
and accelerating particles inside the bubble. 
Part of the TeV emission from HESS J1857$+$026 could be of hadronic origin. Indeed, 
we detect dense molecular material probably associated with the SB that could be acting as a target for
hadronic interactions of protons accelerated inside the bubble.

On the contrary, the old pulsar PSR J1857$+$0300 could also be associated with the SB, although
its contribution to the $\gamma$-ray emission is negligible due to its low energetics. In addition, the associated SNR has already
dissipated into the ambient gas, though it might have contributed in the past to the injection of energy within the SB. 

We conclude that the $\gamma$-ray emission from HESS J1857$+$026 is produced
in the interior of a SB. Very-high-energy photons are probably radiated by a population of relativistic particles 
accelerated in the stellar winds and/or SNRs within the expanding bubble. Furthermore,  
particles could also be accelerated in a PWN powered by PSR J1857$+$0245, which is located inside the SB. 
This study provides evidence that SBs are promising candidates for relevant sources of Galactic cosmic rays.

\section*{Acknowledgments}

A.P. and E.G. are members of the {\sl Carrera del Investigador Cient\'\i fico} of CONICET, Argentina. 
L.D. is postdoctoral fellow of CONICET, Argentina. 
This work was partially supported by Argentina grants awarded by UBA (UBACyT) and ANPCYT.
The National Radio Astronomy Observatory is a facility of the National Science Foundation 
operated under cooperative agreement by Associated Universities, Inc.
This publication makes use of molecular line data from the Boston University-FCRAO Galactic Ring Survey (GRS). 
The GRS is a joint project of Boston University and Five College Radio Astronomy Observatory, funded 
by the National Science Foundation under grants AST-9800334, AST-0098562, AST-0100793, AST-0228993, \& AST-0507657.
This publication makes use of data from FUGIN, FOREST Unbiased Galactic plane Imaging survey with the Nobeyama 45-m telescope, 
a legacy project in the Nobeyama 45-m radio telescope.

\bibliographystyle{aa}  
\bibliography{ref}

\end{document}